\documentclass{epl}

\title{Non-monotonic dependence on disorder in biased diffusion 
on small-world networks}
\shorttitle{Biased diffusion on small-world networks}
\author{Dami\'an H. Zanette} 
\institute{
Consejo Nacional de Investigaciones Cient\'{\i}ficas y T\'ecnicas - 
Centro At\'omico Bariloche, 8400 San Carlos de Bariloche, R\'{\i}o Negro,
Argentina   
}

\pacs{89.75.Hc}{Networks and genealogical trees}
\pacs{05.40.Fb}{Random walks and Levy flights}
\pacs{89.75.Da}{Systems obeying scaling laws} 
                               
\begin{document}

\maketitle

\begin{abstract}
We report numerical simulations of a strongly biased diffusion process on a
one-dimensional substrate with directed shortcuts between randomly chosen
sites, i.e. with a small-world-like  structure. We find that, unlike many
other dynamical phenomena  on small-world networks, this process exhibits
non-monotonic  dependence on the density of shortcuts. Specifically, the 
diffusion time over a finite length is maximal at an  intermediate density.
This density scales with the length  in a nontrivial manner, approaching  zero
as the length grows. Longer diffusion times for intermediate shortcut
densities can be ascribed to the formation of cyclic paths where the 
diffusion process becomes occasionally trapped. 
\end{abstract}

Complex networks are known to underly a wide class of natural  and
artificial systems, ranging from cellular tissues  and human populations to
language structures and computer webs  \cite{WS,Bara,B1,Stan}. Among the
various mathematical models for complex networks, small-world networks 
\cite{WS}
capture a key property of actual biological and social systems, namely, the
fact that sites with distant locations in space may be separated by only a few
links over the network. Small-world networks, in fact, are constructed by
adding randomly distributed shortcuts between distant sites in an initially
ordered lattice, such that the average path length between any two sites 
results in a logarithmic dependence on the network size 
\cite{Barrat,New}. The effect
of the underlying geometry on the dynamics of processes  occurring on
small-world networks has been analyzed with particular emphasis on
propagation phenomena, such as diffusion and  percolation
\cite{p1,p2,Manr,Sato,Kup,rum}. It has been  shown that, in most cases,
any finite density of shortcuts  induces the behavior expected for random
networks, where diffusion is a highly efficient transport mechanism and the
usual linear scaling between mean square displacement and time breaks down. 
On the other  hand, for slightly more complicated dynamics  --where diffusion
is combined with certain reaction-like processes--  the random-network regime
occurs only above a critical (finite)  shortcut density \cite{Kup,rum}.
In all cases, however, the relevant quantities depend monotonically on the
shortcut density. This is also the case for other kinds of phenomena on
small-world networks, such as for Ising dynamics \cite{Barrat}. In this
letter, we introduce a diffusion process on a small-world network with
directed shortcuts where, in contrast, non-monotonic behavior is observed as
the shortcut density is varied --i.e. as the disorder of the 
substrate changes. Our numerical results reveal  well-defined scaling
properties in the limits of low and high  disorder, and a nontrivial scaling
law at the intermediate  regime.

Consider a random walker which moves on a network constructed as follows.
Starting from a one-dimensional array of $N$ sites with nearest-neighbor
connections, a link $L_{ij}$ is established with probability $p$ from each
site $i$ to a randomly chosen site $j$. On the average, $pN$ shortcuts are thus
created --typically, between otherwise distant sites of the array. The
probability $p$ gives the density of shortcuts, and measures the disorder or
{\it  randomness} of the resulting  network. Sites are numbered 
correlatively, $i=1$ to $N$, from one end of the array to  the other. The
random walker starts at the first site, $i=1$.  At each time step, it performs
a forward jump of length $k$,  from site $i$ to site $i+k$. The length $k$ is
drawn at each step from a probability distribution $Q(k)$. If, however, a 
shortcut 
$L_{i+k,j}$ exists between sites $i+k$ and $j$, the final  destination of the
walker in the step under consideration is  site $j$. Otherwise, it stays at
$i+k$. The process stops when  the random walker exits  the system across the
opposite end  of the array, i.e. when from a site $i$ a forward jump of 
length $k>N-i$ takes place.

In this model, consequently, transport is a combination of highly biased,
unidirectional diffusion, whose statistical properties are determined by the
distribution $Q(k)$, and strong mixing due to the shortcuts. The relative
importance of both mechanisms is measured by the probability $p$. Note that,
while the forward motion is here a stochastic process, mixing is
deterministically defined by the links $L_{ij}$, i.e. by the disordered but
frozen structure of the underlying network. Note also that shortcuts are
directed: a link $L_{ij}$ implies deterministic transport from $i$ to $j$,
but not in the opposite direction. The process is reminiscent of certain board
games, such as ``Snakes and Ladders'' and its numerous variations. In this ancient
Hindu game, forward motion represents reincarnation to higher forms of life in
the way towards {\it nirvana}, threatened by the backward shortcuts which
lead to inferior animal forms. In a more physical context, the model may be
interpreted as a diffusion process in a one-dimensional medium under a strong
external field, which induces forward motion, subject to the action of
scattering events along well defined channels --perhaps due to localized
defects or impurities.

The aim of the present analysis is to determine the typical duration $T$ of
the whole transport process as a function of the system length $N$ and the
degree of disorder $p$. In a single realization, the duration $t$ is given by
the number of time steps taken by the random walker to exit the system. The
typical duration $T$ must be defined as a suitable average of $t$ over many
realizations, as explained later. For convenience in the numerical
implementation of the model, we take an exponential distribution of jump
lengths, $Q(k)= q (1-q)^{k-1}$ ($0<q<1$, $k\ge 1$). The average jump length,
$\langle k\rangle =q^{-1}$, gives the corresponding drift velocity, while the
diffusion coefficient is proportional to the mean square dispersion $\langle
k^2\rangle - \langle k\rangle^2 =q^{-2}(1-q)$. For the numerical 
realizations, we choose $q=1/2$.
  
\begin{figure}
\onefigure[width=15cm]{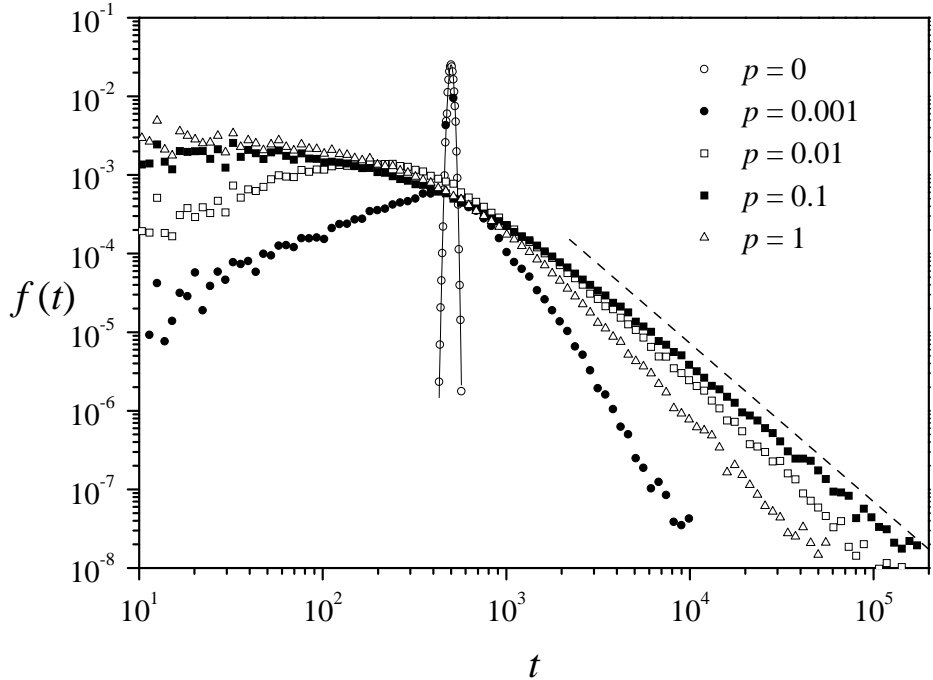}
\caption{Normalized distribution of durations $t$ of the transport process on
a $10^3$-site network, for different values of the  randomness $p$. The
full-line curve corresponds to a Gaussian  fit for $p=0$. The dashed straight
line has slope $-2$. Note  the non-monotonic dependence of the tail slope for
$p>0$.}
\label{f.1}
\end{figure}

We first study the distribution $f(t)$ of the duration $t$ over series of
numerical realizations. It can be easily realized that for large
$N$ and with $p=0$, i.e. in the absence of shortcuts, $f(t)$ is approximately
a Gaussian centered at $qN$ and with width proportional to $q^{-1} (1-q) N$.
Figure \ref{f.1} shows  numerical results for $f(t)$ and several values of
$p$, with $N= 10^3$. Even for very small values of the randomness, $p\sim
N^{-1}$ --where, on the average over realizations of the network, only one 
shortcut is added-- the distribution develops a wide background over a large
range of values of $t$. As $p$ grows, this background  rapidly replaces the
Gaussian peak and widens, while a  power-law  large-$t$ tail becomes apparent.
For $p\approx 0.1$ the slope of this power-law tail attains its minimum
(absolute) value, equal to $-2$ within numerical precision, and the
distribution exhibits its broadest shape. For larger randomness, the  tail
recedes and  becomes steeper. At $p=1$, the slope has grown to approximately 
$-2.6$. This is a first indication of non-monotonic behavior as a function of
the randomness.

A more compact description of the dependence of $f(t)$ with $p$ is given by
the typical time $T$ taken by the random walker to exit the system. Such
typical time, however, must be defined with caution. Indeed, a naive
definition in terms of the average  $\langle t \rangle = \int t f(t) \upd 
t$ could
fail to give numerical  results with acceptable precision for the values $p$
where the  slope of the large-$t$ tail of $f(t)$ is close to $-2$. For such
randomness, large fluctuations between single realizations are  expected in
the value of $\langle t \rangle$. If the slope does  in fact reach $-2$, the
average $\langle t \rangle$  would directly be ill-defined. Consequently, we
define
\begin{equation} \label{T}
T = \left[ \int_0^\infty  t^\alpha f(t) \upd t \right]^{1/\alpha},
\end{equation}
with $0<\alpha<1$. As far as $0<\alpha<1$, the choice is arbitrary.
Below, we take $\alpha=1/2$.  

\begin{figure}
\onefigure[width=10.5cm]{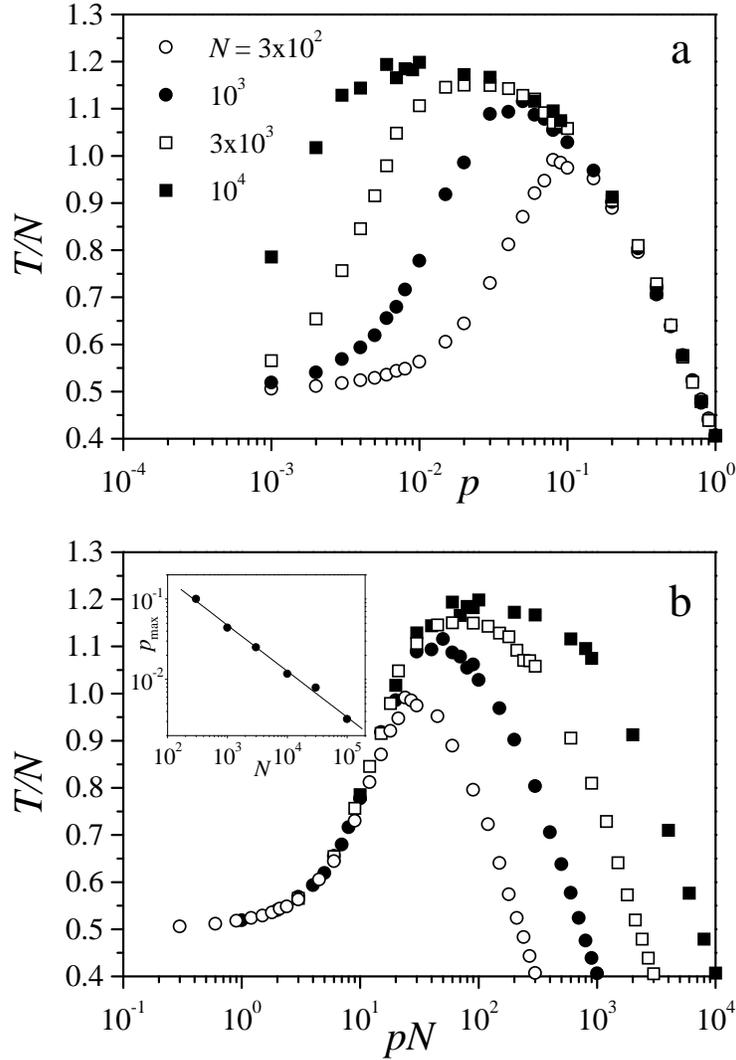}
\caption{Typical duration of the transport process --defined as in eq. 
(\ref{T}) with $\alpha=1/2$-- as a function of the randomness $p$, for
several values of the length $N$. (a) $T/N$ versus $p$, (b) $T/N$  versus
$pN$. Symbols are the same for both plots. The insert shows  the position of
the maximum, $p_{\max}$, as a function of $N$.}
\label{f.2}
\end{figure}

Figure \ref{f.2} shows the typical time $T$ as a function of randomness, for
several system lengths ranging from $N=3\times  10^2$ to $10^4$. Similar
results, not shown in the figure,  were obtained for up to $N=10^5$.
Non-monotonicity is apparent. While for $p\to 0$ we find the expected result
$T\approx qN=N/2$  and for $p\to 1$ we have $T\approx 0.4N$, for intermediate 
randomness the ratio $T/N$ reaches considerably higher values. As we discuss
in more detail below, the randomness $p_{\max}$ at which $T$ attains it
maximum decreases with $N$,  and the maximum itself grows. The two plots in
fig. \ref{f.2} reveal  two well-defined scalings in the dependence of $T$ on
$p$.  For $p\approx 1$ we have (see fig. \ref{f.2}a)
\begin{equation} \label{s1}
T=NF_1(p),
\end{equation}
while for $p\approx 0$, we find  (see fig. \ref{f.2}b)
\begin{equation} \label{s0}
T=NF_2(pN).
\end{equation}
The scaling functions $F_1$ and $F_2$ are numerically found to be 
non-singular at the respective limits. 

The scaling properties reported in eqs. (\ref{s1}) and  (\ref{s0}) can be
explained in qualitative terms with simple arguments. For sufficiently small
randomness, only a few  shortcuts are encountered in each realization of the
process. Since shortcut ends are distributed at random, the average shortcut
length is of order $N$. Therefore, the average duration $qN$ at $p=0$ is
modified by a quantity of order  $N$ as $p$ grows. Note that forward and
backward shortcuts  respectively diminish and increase the value of $T$. In
any  case, the resulting typical duration is proportional to $N$. Now, the
number of different sites visited during a realization is also of order $N$,
so that the probability of finding a shortcut is proportional to $pN$.
Consequently, the typical duration $T$ is expected to depend on $p$ through
the product  $pN$ only, as in eq. (\ref{s0}). Finally, since forward and
backward shortcuts are equally probable, why does $T$  increase with $p$? To
answer this question, imagine a  realization of the network with a single
shortcut starting at site $i$. If the shortcut is forward, $L_{ij}$ with
$i<j$, it will be
encountered by the random walker at most once in each realization, if  the
site $i$ is visited. On the other hand, if the shortcut is backward, $L_{ij}$ 
with $i>j$, there is a
finite probability that the walker visits site $i$ several times, remaining
temporarily  ``trapped'' into a cycle until it succeeds at skipping site 
$i$. Thus, the contribution of each backward shortcut to the  typical
duration $T$ is effectively more important than the  opposite contribution of
each forward shortcut.

When the density of shortcuts is large, $p\approx 1$, there is a high
probability for the random walker to encounter a  shortcut at each step and
therefore land at a random site  in the network. The probability of exiting
the system  becomes large only when landing in the final end of the  array,
$i\approx N$. From there, diffusion can transport  the walker outside the
system, thus ending the process.  The size of the region where the probability
of abandoning the system by diffusion is above any fixed threshold is
independent of $N$, so that the probability of reaching that region at any
time step is of order $N^{-1}$. The typical duration of the process is
therefore proportional  to $N$. As for the dependence on $p$, $T$ decreases as
$p$ grows simply because the probability of finding a shortcut leading to the
final region of the array increases  proportionally to the randomness. Note
that the total  number of shortcuts leading to (or starting from) a
region of finite size is proportional to $p$ and  independent on $N$.

Since in the two regions $p\approx 0$ and $p\approx 1$  the scaling regimes
of $T$ as a function of $p$ and $N$  are different, the scaling properties in
the intermediate  region --where the duration is maximal--  are not 
unambiguously defined. We have therefore recorded the randomness $p_{\max}$
at which the typical duration attains its maximum $T_{\max}$, for several
system sizes up to $N= 10^5$. The insert of fig. \ref{f.2}b shows $p_{\max}$
as a  function of $N$. Over the whole range, the position of the maximum is
well approximated by a power law $p_{\max}  \sim N^{-\gamma}$, with a
nontrivial scaling exponent 
\begin{equation}
\gamma=0.57 \pm 0.02.
\end{equation}
The maximum duration $T_{\max}$, on the other hand, does not show a
well-defined behavior as a function of $N$.  Our numerical precision does not
make it possible to  discern between a saturation towards $T_{\max} /N 
\approx 1.2$ or a slight growth of the ratio $T_{\max}  /N$ with length.
Discerning between these two possibilities would require realizations for
substantially larger values of $N$.

In the intermediate regime, the effect of temporary trapping into cycles
which contain backward shortcuts  --discussed above in connection with the
regime of  small randomness-- becomes dominant. Due to the fact  that the
probability of skipping any given  site in  the lattice is finite, the random
walker cannot be  indefinitely trapped in one of such cycles. We find 
however that repetition of the same cycle up to four or five times in a
single realization is quite a  common occurrence, which explains the
relatively  large durations observed in this regime. The role  of cycles in
the dynamics explains also the power-law tail in the distribution of durations
shown in  fig. \ref{f.1}. In fact, it is known from other (deterministic)
transport processes on lattices with sites connected by randomly distributed 
links --just like the shortcuts in our system-- that cycle lengths are
distributed according to a power  law \cite{ham}. It is interesting to point
out   that the power-law distribution is observed even  for $p=1$ (see fig.
\ref{f.1}), which suggests  that in that limit the process is  still governed 
by the existence of cycles. This complex ingredient prevents the analytical
calculation of $T$ for $p\approx 1$. In fact, a naive calculation of  $f(t)$
using the arguments given above to explain  the dependence of $T$ on $p$ and
$N$ in that limit, predicts  an exponential distribution.

In summary, we have presented and numerically studied  a stochastic transport
process on a small-world-like  one-dimensional substrate, which exhibits 
non-monotonic dependence on the substrate randomness $p$. This non-monotonicity 
is the result of competition  between the formation of trapping cycles as the
randomness grows and the overall improvement of transport efficiency due to
the presence of shortcuts. Non-monotonicity naturally defines two separated
regimes in the randomness domain, which are found to obey well-differentiated
scaling laws in terms of $p$ and the system size $N$. As for many other
dynamical processes on small-world networks \cite{Barrat},  the 
small-randomness regime is observed below  $p\sim N^{-1}$ and thus 
collapses in a trivial way as $N$
increases. In our system, however, we find  a third, intermediate regime
--where the process  becomes extremal and non-monotonicity is therefore 
realized-- characterized by a nontrivial scaling law, $p\sim N^{-\gamma}$
with $\gamma \approx 0.6$. 

It is interesting to point out that, to our knowledge, the only previous 
report of non-monotonic dependence on 
randomness for dynamical processes on small-world networks also involves 
a directed network \cite{San}. This leads to conjecture that this kind of 
non-monotonicity may be a common feature in directed small-world networks. 
Directed shortcuts in small-world networks --and,  in general, in complex 
networks-- have  been scarcely treated in the literature, in spite of the 
fact that they should be essential to the mathematical description of certain  
non-physical (biological, social) interactions that have recently attracted 
much attention. Our analysis suggests that such kind of links can add 
considerable complexity to the involved processes.

\acknowledgments
This work was partially carried out at the Abdus Salam 
International Centre for Theoretical Physics (Trieste, Italy).

\end{document}